\documentclass[slac_one]{revtex4}
\usepackage{graphicx}
\usepackage{fancyhdr}
\begin{document}

\title{{\small{Hadron Collider Physics Symposium (HCP2008),
Galena, Illinois, USA}}\\ 
\vspace{12pt}
Diboson Production at the Tevatron} 

%

\author{Ia Iashvili}
\affiliation{SUNY, Buffalo, NY 14260, USA}

\begin{abstract}
We present the latest results on the production
of  $WW$, $WZ$, $W\gamma$, $Z\gamma$
and $ZZ$ events at the Fermilab Tevatron Collider.
The results are based on the analyses
of  0.2 -- 2~fb$^{-1}$ of data collected in $p\bar{p}$ collisions
at $\sqrt{s}=1.96$~TeV
by CDF and 
D\O\ experiments during the Tevatron Run~II. 
Analyses of the diboson production processes 
provide crucial test of the Standard Model,
directly probing its predictions on the Trilinear Gauge 
Couplings.
\end{abstract}

\maketitle

\thispagestyle{fancy}

\section{INTRODUCTION} 

The Standard Model (SM) makes precise predictions for the couplings
between gauge bosons thanks to the non-abelian nature of its
$SU(2)_L \times U(1)_Y$ symmetry. 
These self-interactions are described by the 
trilinear $WW\gamma$,
$WWZ$, $Z\gamma\gamma$ and $ZZ\gamma$ and $ZZZ$ gouge couplings (TGCs),
which can be directly tested in the pair productions of the gauge bosons.
Therefore datasets of
$WW$, $W\gamma$, $Z\gamma$, $WZ$ and $ZZ$ candidate
events produced in $p\bar{p}$ collisions at $\sqrt{s}=$1.96~TeV
at the Tevatron $p\bar{p}$ Collider
provide crucial testing ground for  SM. 
Any deviations from the SM predictions can indicate presence of
New Physics.
Furthermore, diboson processes have signatures similar to
that of the Higgs production at the Tevatron, 
and constitute background to the Higgs searches. Thus 
detailed understanding of the diboson processes at the
Tevatron is viewed as a first step towards
probing the Higgs boson production.

Production cross sections for the diboson processes at the Tevatron
are a few orders of magnitude smaller compared 
to that of the inclusive $W$ and $Z$
productions. Diboson events with the leptonic decays of $W$ 
and $Z$ bosons provide final states with the lowest background contamination,
but also suffer from small branching ratios.

The Tevatron Collider has already delivered more than 4~fb$^{-1}$
of data to the CDF and D\O\ experiments.
These large datasets allow to probe the  diboson
processes even with very small production cross section times
branching fraction, of the order
of a few femtobarns. The results presented here are
based on 0.2 -- 2~fb$^-1$ of data.

\section{$WW\to\ell\ell\nu\nu$ production}

Pair production of the $W$ bosons at the Tevatron, 
$p\bar{p} \to W^+W^-$,
proceeds through $Z/\gamma$ exchange. Thus $WW$ events
allow to probe trilinear $WWZ/WW\gamma$ couplings.
Furthermore, $W^+W^-$ events are dominant irreducible background
to the Higgs searches in the $H\to W^+W^-$ channel,
and their understanding is important.
When followed by leptonic decays,
$W\to\ell\nu$ ($\ell=e$ or $\mu$), of both $W$s, $WW$ production leads to 
final states with two high-$p_T$ isolated leptons of opposite
sign, $e^{\pm}e^{\mp}$, $\mu^{\pm}\mu^{\mp}$ or $e^{\pm}\mu^{\mp}$,
and large transverse missing energy, $E_T^{miss}$, due 
to escaping neutrinos. 
There are many other SM processes which can give
the similar event signature: 
$W(\to\ell\nu)$+jets production with a jet
faking electron or containing muon,  
$Z/\gamma^*\to\ell\ell$, $t\bar{t}$, $WZ$ and $ZZ$ processes
can all contribute to the background.
Signal separation from the background is achieved
by rejecting $e^+e^-$ and $\mu^+\mu^-$ events with dilepton mass
consistent to $M_Z$, by vetoing large hadronic activities,
and removing events where missing $E_T$ is likely to have
originated from jet mis-measurements.
The $WW$ production signal has been established by
both, D\O\ and
CDF Collaborations with already $\simeq$240~pb$^{-1}$~\cite{ww_xs_d0}
and $\simeq$~200~pb$^{-1}$~\cite{ww_xs_cdf_published} of data, respectively. 
The measured cross sections of 
$\sigma(WW)=13.8^{+4.3}_{-3.8}$~(stat)~$^{+1.2}_{-0.9}$~(syst)~$\pm 0.9$~(lumi)~pb~\cite{ww_xs_d0} by D\O, and
$\sigma(WW)=13.6 \pm 2.3$~(stat)$\pm 1.6$~(syst)~$\pm 1.2$~(lumi)~pb  
by CDF using ${\cal{L}}\simeq 825$~pb$^{-1}$ of data~\cite{ww_xs_cdf_latest},
are in agreement with
a SM Next-to-Leading Order (NLO) prediction of 
12.0 -- 13.5~pb~\cite{ww_xs_theory,xs_theory}.

\begin{figure*}[t]
\centering
\includegraphics[width=70mm]{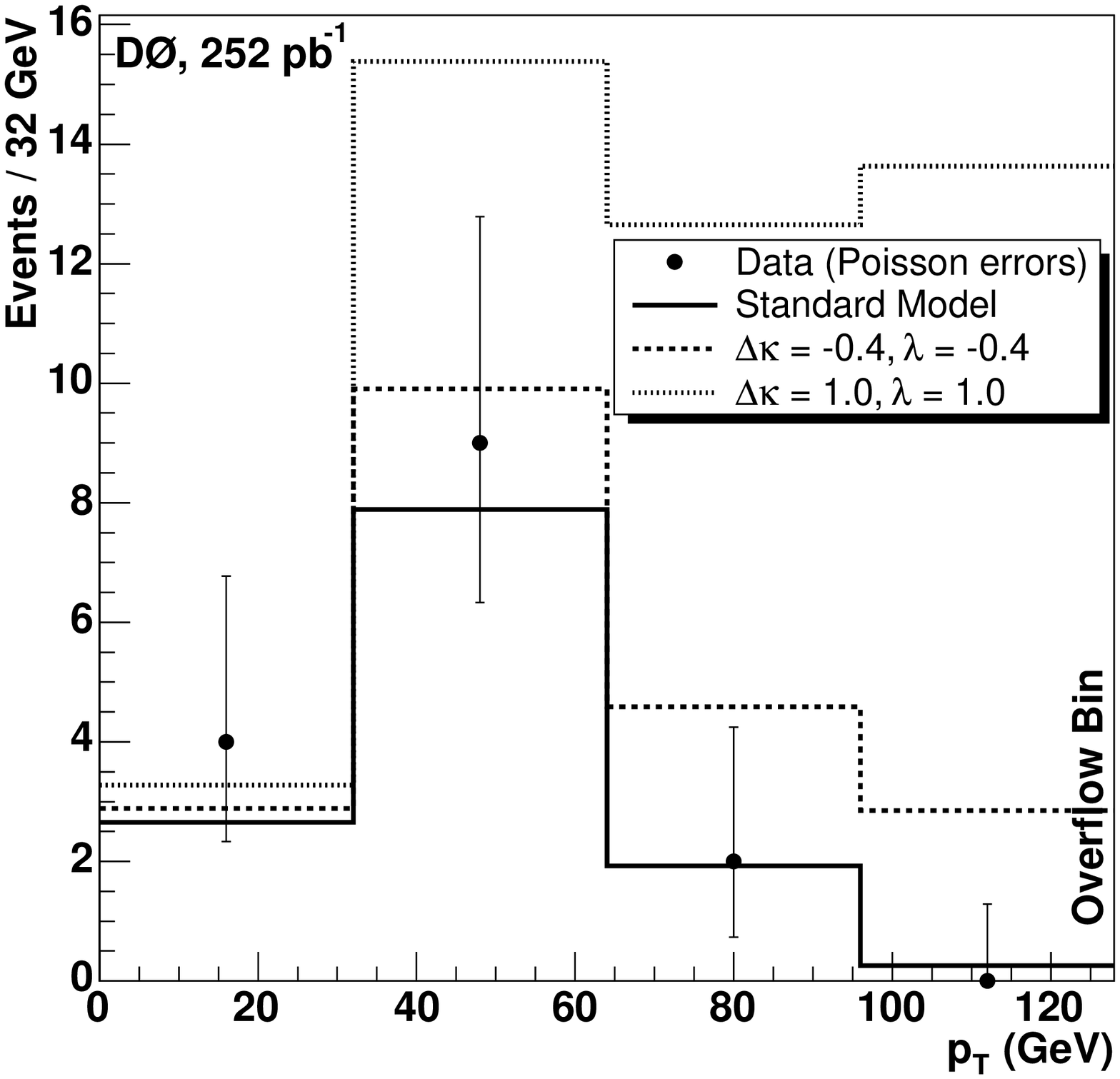}\hspace*{15mm}
\includegraphics[width=73mm]{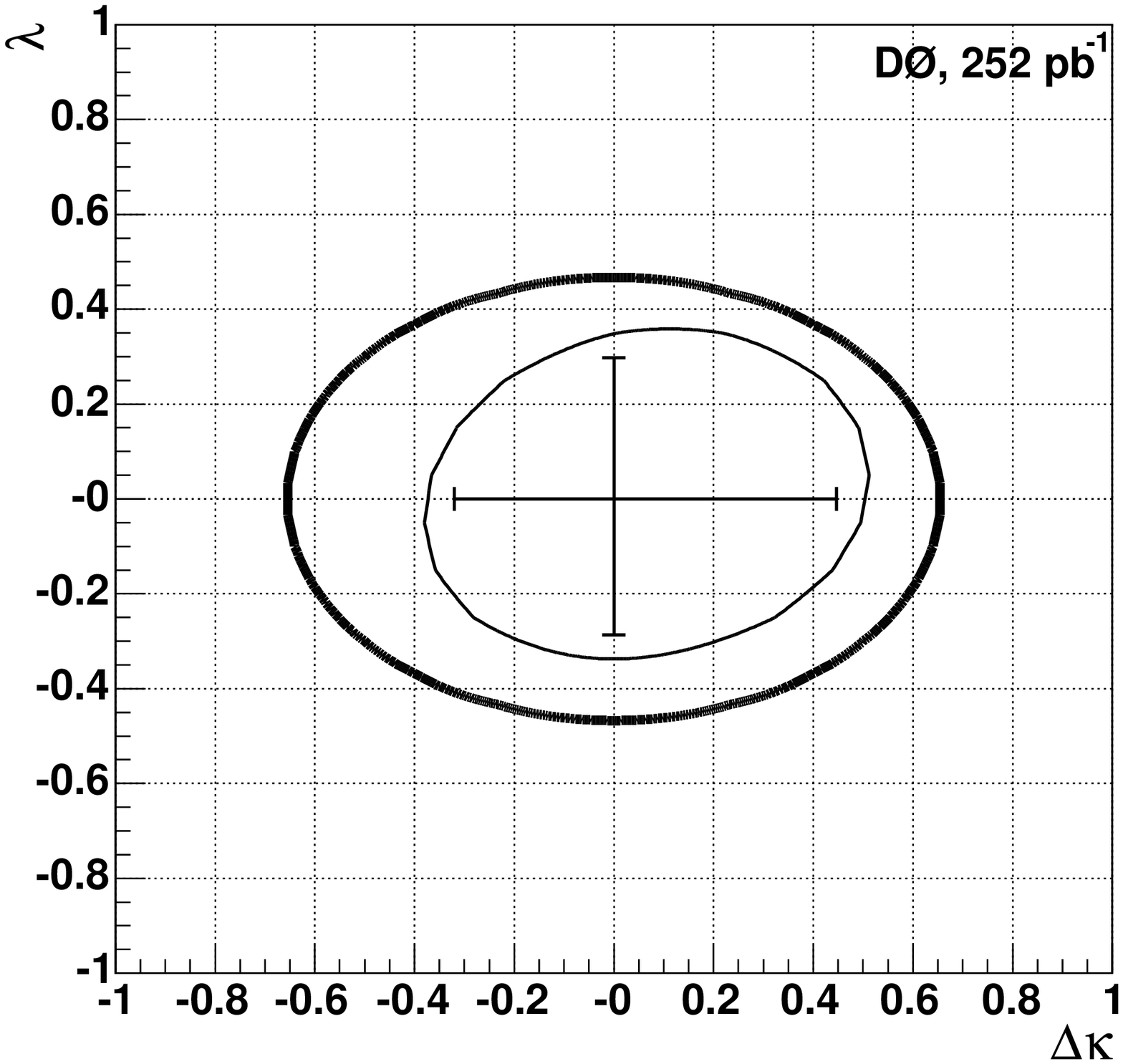}
\caption{Left: Distribution of the leading lepton $p_T$ for 
$WW\to e^{\pm} \mu^{\mp}$ candidates in D\O \ data, 
and expectations from the SM
(solid line) and two anomalous coupling scenarios (dashed lines)~\cite{ww_ac_d0}.
Right: One- (ticks along the axes) and two-dimensional (the inner
curve) 95\% C.L. limits at $\Lambda =2.0$~TeV assuming equal $WWZ$ and 
$WW\gamma$ couplings. The bold curve is the unitarity 
limit. The limits are
obtained from D\O\  $WW\to \ell^{\pm} \ell^{\mp}$ analysis~\cite{ww_ac_d0}.} 
\label{fig:ww_ac}
\end{figure*}

The general Lorentz invariant effective Lagrangian describing
$WWV$ ($V=\gamma$ or $Z$) vertices~\cite{112,113}
has seven parameters for each of the $WW\gamma$ and $WWZ$ vertices. With the
assumption of electromagnetic gauge invariance and C and P conservation, the number of independent couplings is reduced to five, and the Lagrangian takes the form:
\begin{equation}
\frac{{\cal{L}}_{WWV}}{g_{WWV}} = 
ig_1^V(W_{\mu\nu}^{\dagger}W^{\mu}V^{\nu} - W_{\mu}^{\dagger}V_{\nu}W^{\mu\nu})
+ i\kappa_V W_{\mu}^{\dagger}W_{\nu}V^{\mu\nu} + \frac{i\lambda_V}{M_W^2}W_{\lambda\mu}^{\dagger}W^{\mu}_{\nu}V^{\nu\lambda}
\end{equation}
where $W^{\mu}$ is the $W^-$ field, $W_{\mu\nu}=\delta_{\mu}W_{\nu}-\delta_{\nu}W_{\mu}$,
$V_{\mu\nu}=\delta_{\mu}V_{\nu}-\delta_{\nu}V_{\mu}$, and $g_1^{\gamma}=1$. The
overall couplings are $g_{WW\gamma}=-e$ and $g_{WWZ}=-e~cot \theta_W$. 
The five remaining parameters are $g_1^Z$, $\kappa_Z$, $\kappa_{\gamma}$, $\lambda_Z$, 
and $\lambda_{\gamma}$. In the SM, $g_1^Z = \kappa_Z = \kappa_{\gamma}=1$ and
$\lambda_Z = \lambda_{\gamma} = 0.$ The couplings $g_1^Z$, $\kappa_Z(\gamma)$
are often written in terms of their deviation from the SM values as
$\Delta g_1^Z =g_1^Z - 1$, $\Delta \kappa_{Z(\gamma)} = \kappa_{Z(\gamma)} - 1$.

\begin{table}[b]
\begin{center}
\caption{One-dimensional 95 \% C.L. limits with various
assumptions relating the $WW\gamma$ and $WWZ$ couplings and 
various values of form factor scale $\Lambda$. Parameters 
that are not constrained  by the coupling relationships 
are set to their SM values. The limits are obtained in D\O\ WW 
analysis~\cite{ww_ac_d0}}
\vspace*{2mm}
\begin{tabular}{|c|c|c|c|c|c|}
\hline 
\textbf{Assumptions on}  & \textbf{$WW\gamma=WWZ$} &
\textbf{$WW\gamma=WWZ$} &  \textbf{$HISZ$} &  \textbf{$SM WW\gamma$} &  
\textbf{SM $WWZ$}
\\
\textbf{couplings and $\Lambda$}  & \textbf{$\Lambda=$1.5~TeV}  & \textbf{$\Lambda=$2.0~TeV}& \textbf{$\Lambda=$1.5~TeV}&\textbf{$\Lambda=$2.0~TeV} &\textbf{$\Lambda=$1.0~TeV}
\\ \hline 
95\% C.L. & -0.31 $<\lambda <$ 0.33 &
            -0.29 $<\lambda <$ 0.30 &
            -0.34 $<\lambda <$ 0.35 &
            -0.39 $<\lambda_Z <$ 0.39 &
            -0.97 $<\lambda_{\gamma} <$ 1.04     \\
Limits     & -0.36 $<\Delta\kappa <$ 0.47  &
             -0.32 $< \Delta\kappa <$ 0.45 &
             -0.57 $< \Delta\kappa_{\gamma} <$ 0.75 &
             -0.45 $< \Delta\kappa_Z <$ 0.55 &
             -1.05 $< \Delta\kappa_{\gamma}<$ 1.29 \\
\hline
\end{tabular}
\label{table:ww_ac}
\end{center}
\end{table}

One effect of introducing anomalous coupling parameters 
into the SM Lagrangian is 
an increase of the cross section for the $q\bar{q}\to Z/\gamma \to W^+ W^-$ production  with increasing parton center-of-mass energy $\sqrt{\hat{s}}$. 
To keep the cross section from diverging, the anomalous 
coupling must vanish as $s\to \infty$. This is achieved by introducing a dipole form 
factor for arbitrary coupling $\alpha$ 
($g_1^Z$, $\kappa_Z$, $\kappa_{\gamma}$, $\lambda_Z$  or $\lambda_{\gamma}$):
$\alpha(\hat{s}) =\frac{\alpha_0}{(1+\frac{\hat{s}}{\Lambda^2})^2}$,
where the form factor $\Lambda$ is set by new physics. For a given
value of $\Lambda$, there is an upper limit on the size of
the coupling, beyond which unitarity is exceeded.

Non-SM couplings enhance $WW$ production cross section,
particularly at high values of the boson $p_T$. To probe
$WWZ/WW\gamma$ TGCs, observed $p_T$ spectrum of the two leptons are 
fitted to the templates of the $WW$ MC events produced
for scanned values of the non-SM couplings.
Figure~\ref{fig:ww_ac} (left) shows distribution of the leading lepton
$p_T$ in $WW\to e^{\pm} \mu^{\mp}$ candidate events together
with the expected distributions from the SM, and for
the two representative values of the non-SM couplings~\cite{ww_ac_d0}.
Figure~\ref{fig:ww_ac} (right) shows one- and two-dimensional 95\% C.L.
limits on $\Delta\kappa$ and $\lambda$ parameters
at $\Lambda$=2.0~TeV. The limits 
are derived under the assumption of equal 
$WWZ$ and $WW\gamma$ couplings. 
Table~\ref{table:ww_ac} summarizes obtained limits on anomalous
$WWZ$ and $WW\gamma$ couplings for various values of 
$\Lambda$ parameter and for four different assumptions
on anomalous coupling interrelations. In the first relationship,
the $WW\gamma$ and $WWZ$ parameters are equal; the second 
relationship, the HISZ parametrization~\cite{hisz} imposes
$SU(2)\times U(1)$ symmetry upon the coupling
parameters; for the two other relationships, either the SM
$WW\gamma$ or $WWZ$ interaction is fixed, while the other
parameters are allowed to vary. In all cases, parameters
which are not constrained by the coupling relationships are
set to their SM values.

\begin{figure*}[t]
\centering
\includegraphics[width=78mm]{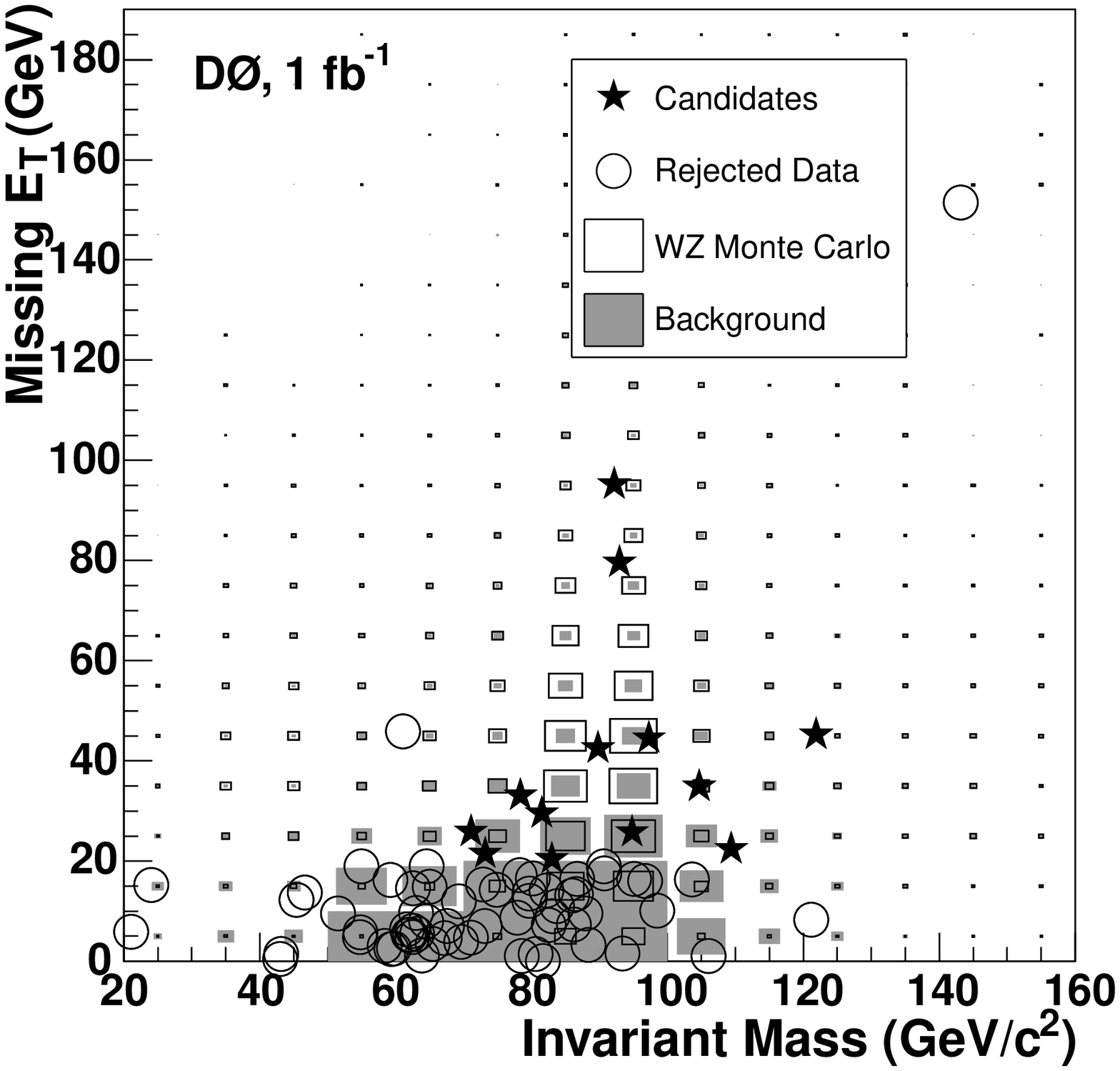}\hspace*{8mm}
\includegraphics[width=78mm]{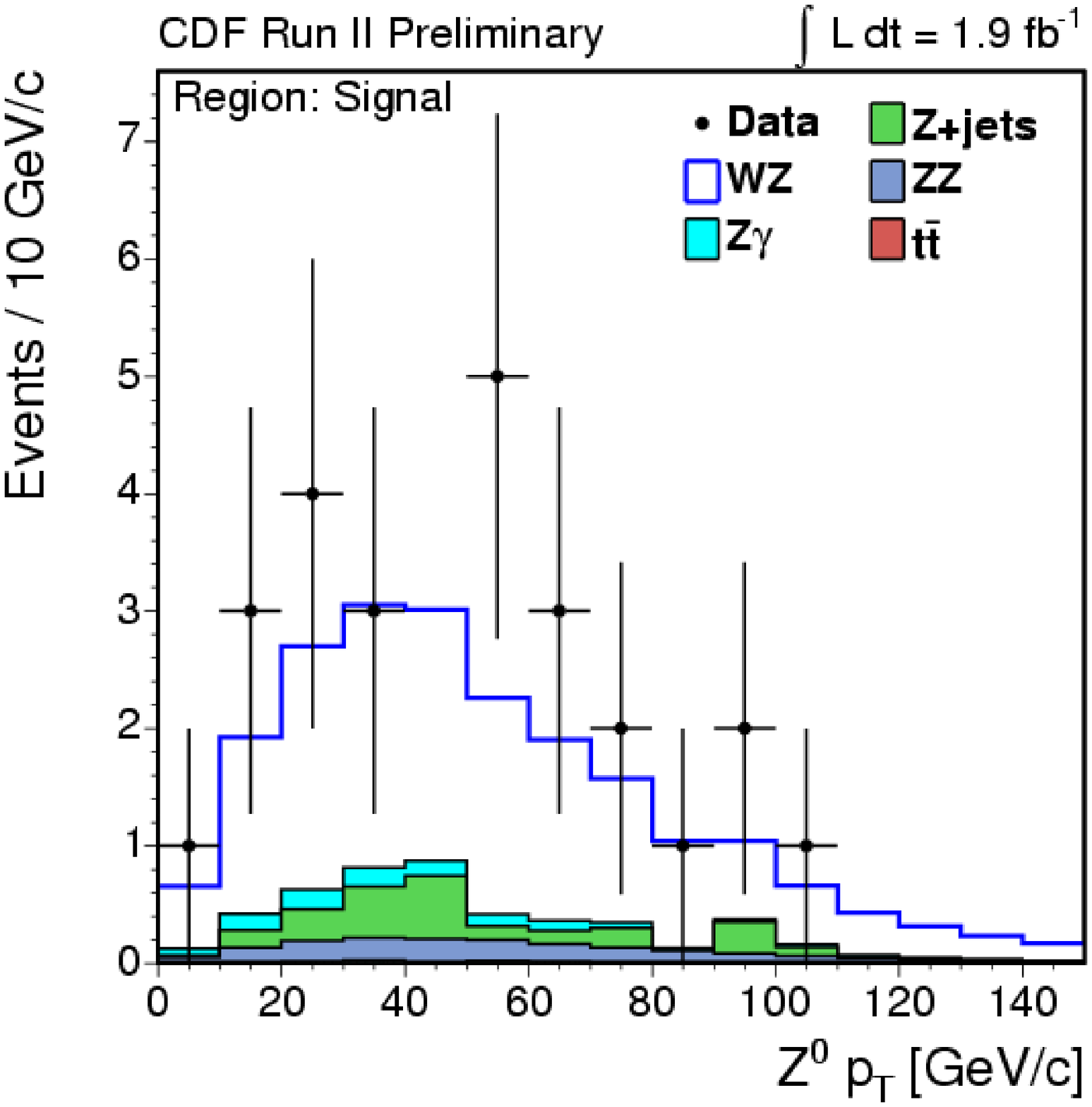}
\caption{Left: Missing $E_T$ versus dilepton invariant mass for D\O\
 $WZ\to\ell'\ell\ell\nu$ candidate events~\cite{wz_d0}. The open boxes 
represent the expected $WZ$ signal. The gray boxes represent the sum of
the estimated backgrounds. The black stars are the data 
that survive all selection  criteria. The open circles
are data that fail either the dilepton invariant mass criterion or
have low missing $E_T$. Right: Distributions of 
$Z$ $p_T$ for CDF $WZ\to\ell'\ell\ell\nu$ candidate
events (points), for various expected background processes (hatched
histograms) and for signal+background (open histograms)~\cite{wz_ac_cdf_preliminary}.
} \label{fig:wz}
\end{figure*}

\section{$WZ\to\ell'\ell\ell\nu$ production}

$WZ$ production, when accompanied with leptonic decays of
both bosons, $W\to\ell\nu$,
$Z\to\ell\ell$, gives very distinct experimental signature.
The final states contain three high-$p_T$ isolated
leptons, of which at least two have the same flavor and 
the invariant mass consistent with $M_Z$, and large missing $E_T$
and transverse mass $M_T(\ell, E_T^miss)$. 
Backgrounds arise from $Z+$jets, $Z\gamma$, $ZZ$ and $t\bar{t}$ productions.

CDF and D\O\ Collaborations have both studied $WZ$ production~\cite{wz_d0_first,wz_d0,wz_cdf,wz_cdf_preliminary,wz_ac_cdf_preliminary}.
The signal has been established at more than 5$\sigma$ statistical 
level by CDF with 1.1~fb$^{-1}$ of data~\cite{wz_cdf}. Measured cross sections
are $\sigma(WZ)=2.7^{+1.7}_{-1.3}$~pb by D\O\ with
${\cal{L}}=1.0$~fb$^{-1}$~\cite{wz_d0},  and $\sigma(WZ)=
4.3^{+1.3}_{-1.0}$~(stat)~$\pm0.2$~(syst)$\pm0.3$~(lumi)~pb by CDF with
${\cal{L}}=1.9$~fb$^{-1}$~\cite{wz_cdf_preliminary}.
These agree with the SM NLO prediction of
$\sigma(WZ)=3.68\pm0.25$~pb~\cite{xs_theory}.
Figure~\ref{fig:wz} (left) shows distribution of
$E_T^{miss}$ versus dilepton invariant mass
in D\O\ $WZ\to\ell'\ell\ell\nu$ candidate events together with 
the expected $WZ$ signal and the estimated backgrounds.

The Fermilab Tevatron currently is the  only particle accelerator that 
can produce the charged state $WZ$.  The $WZ$ events  provide
a unique opportunity to study the
$WWZ$ TGCs without any assumption on the values of the $WW\gamma$ couplings.
As discussed in the previous section, measurements of TGCs using the 
$WW$ events are sensitive to
both the $WW\gamma$ and $WWZ$ couplings at the same time and must make some
assumption as to how they are related to each other.
Non-SM  anomalous TGCs will enhancement the $WZ$
production cross section, and modify the 
shapes of kinematic distributions, such as the $W$ and $Z$ bosons
transverse momenta. By comparing the measured cross section and
$P_T^Z$ distribution to the SM prediction and to  models with 
anomalous TGCs, the Tevatron experiments
set limits on the three coupling parameters: $\lambda_Z$, $\Delta g_1^Z$,
and $\Delta\kappa_Z$. A comparison of the observed $Z$ boson 
$p_T$ distribution in CDF data with SM  predictions for signal
and background is shown
in Fig.~\ref{fig:wz} (right). The table~\ref{table:wz_ac}
summarizes obtained 95\% C.L. limits on the coupling parameters
for the scale factor $\Lambda=2$~TeV.

\begin{table}[t]
\begin{center}
\caption{One-dimensional 95 \% C.L. limits 
obtained by D\O\ and CDF in  $WWZ$ analysis.
The limits correspond to form factor scale of $\Lambda=2$~TeV.}
\vspace*{2mm}
\begin{tabular}{|c|c|}
\hline
\textbf{D\O, ${\cal{L}}=$1.1~fb$^{-1}$~\cite{wz_d0}} & 
\textbf{CDF, ${\cal{L}}=$1.9~fb$^{-1}$~\cite{wz_ac_cdf_preliminary}} \\ \hline
-0.17 $ < \lambda_Z < $ 0.21    &   -0.13 $<\lambda_Z<$ 0.14 \\ \hline
-0.14 $ <  \Delta g_Z <$ 0.34   &   -0.13 $ <  \Delta g_Z <$ 0.23 \\ \hline
-0.12 $ <  \Delta\kappa_Z=\Delta g_Z <$ 0.29   &
-0.76 $ <  \Delta\kappa_Z=\Delta g_Z <$ 1.18   \\ \hline
\end{tabular}
\label{table:wz_ac}
\end{center}
\end{table}

\begin{figure*}[b]
\centering
\includegraphics[width=85mm]{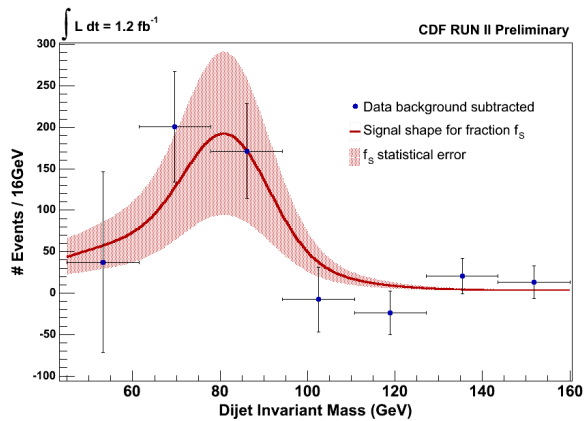}
\caption{Background subtracted distribution of dijet invariant mass
in CDF $WW/WZ\to\ell\nu jj$ candidate events~\cite{ww_wz_jj_cdf}.} 
\label{fig:Fit_Result_3}
\end{figure*}

\section{$WW/WZ\to \ell\nu j j$ production}

The CDF Collaboration has also searched $WW/WZ$ production in the $\ell\nu j j$
final state~\cite{ww_wz_jj_cdf}. The signature arises when
W decays leptonically, $W\to\ell\nu$, and
the associated boson decays hadronically, $W/Z\to jj$. 
The resulting final state is similar to that of  Higgs 
production in $WH\to \ell\nu b\bar{b}$ channel, and is 
experimentally much more challenging than the fully leptonic 
decay modes of $WW$ and $WZ$ productions. The background
arises due to $W/Z$+jets, QCD multijet, $t\bar{t}$ events.
After selecting events with a 
high-$p_T$ lepton, large missing $E_T$ and transverse mass
$M_T(l, E_T^miss)$, and $\ge$2 jets, signal/background ratio
is less than $1\%$. Several discriminating kinematic variables
are combined into Neural Net to achieve further
separation of the signal from the background. Finally, the signal
is extracted by fitting observed dijet mass $M(jj)$ distribution
to the templates of the expected signal and background distributions.

Figure~\ref{fig:Fit_Result_3} shows distribution of the dijet invariant mass
in the candidate events after subtracting background
contribution. Measured cross section times branching ratio is
$\sigma \times BR =1.47 \pm 0.77$~(stat)$\pm 0.38$~(syst)~pb. 
Since observed signal has less than 3$\sigma$ statistical significance, 
95 \% C.L. limit is also set on the cross section times branching ratio:
$\sigma \times BR < 2.88 $~pb.
The results are in agreement with the theory calculations of
 $\sigma \times BR =2.09 \pm 0.14$~pb~\cite{xs_theory}.

\section{$W\gamma\to\ell\nu\gamma$ production and study of Radiation Amplitude
Zero}

Production of $W\gamma$ events at the Tevatron is
studied in the leptonic decay mode of $W\to\ell\nu$
 which leads to the final state containing lepton, neutrino
and a photon.
The events are selected by requiring
a high $p_T$ lepton, large values of $E_T^{miss}$ and transverse
$M_T(\ell, E_T^{miss})$, and a photon with $E_T$ above 7 or 8~GeV. The
dominant background arises from $W$+jets production where a 
jet mimics a photon.
Inclusive $Z\to\ell\ell$ production and $Z\gamma$ events 
can also contribute to the background.
Both, D\O\ and CDF
Collaborations have measured $W\gamma$ production cross 
section~\cite{wg_d0, wg_cdf} and found good agreement with the SM
expectation.
 
\begin{figure*}[t]
\centering
\includegraphics[width=80mm]{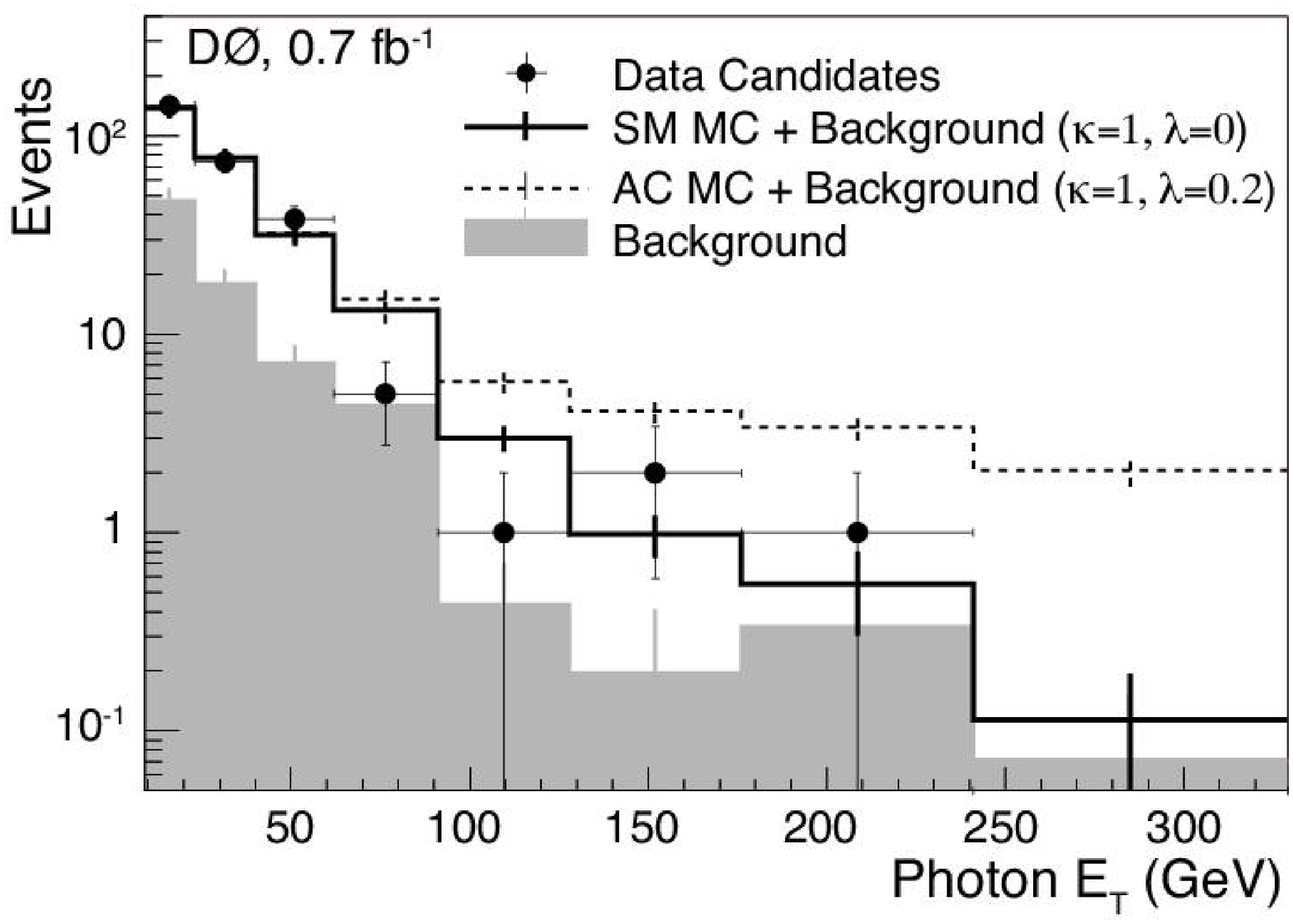}\hspace*{9mm}
\includegraphics[width=80mm]{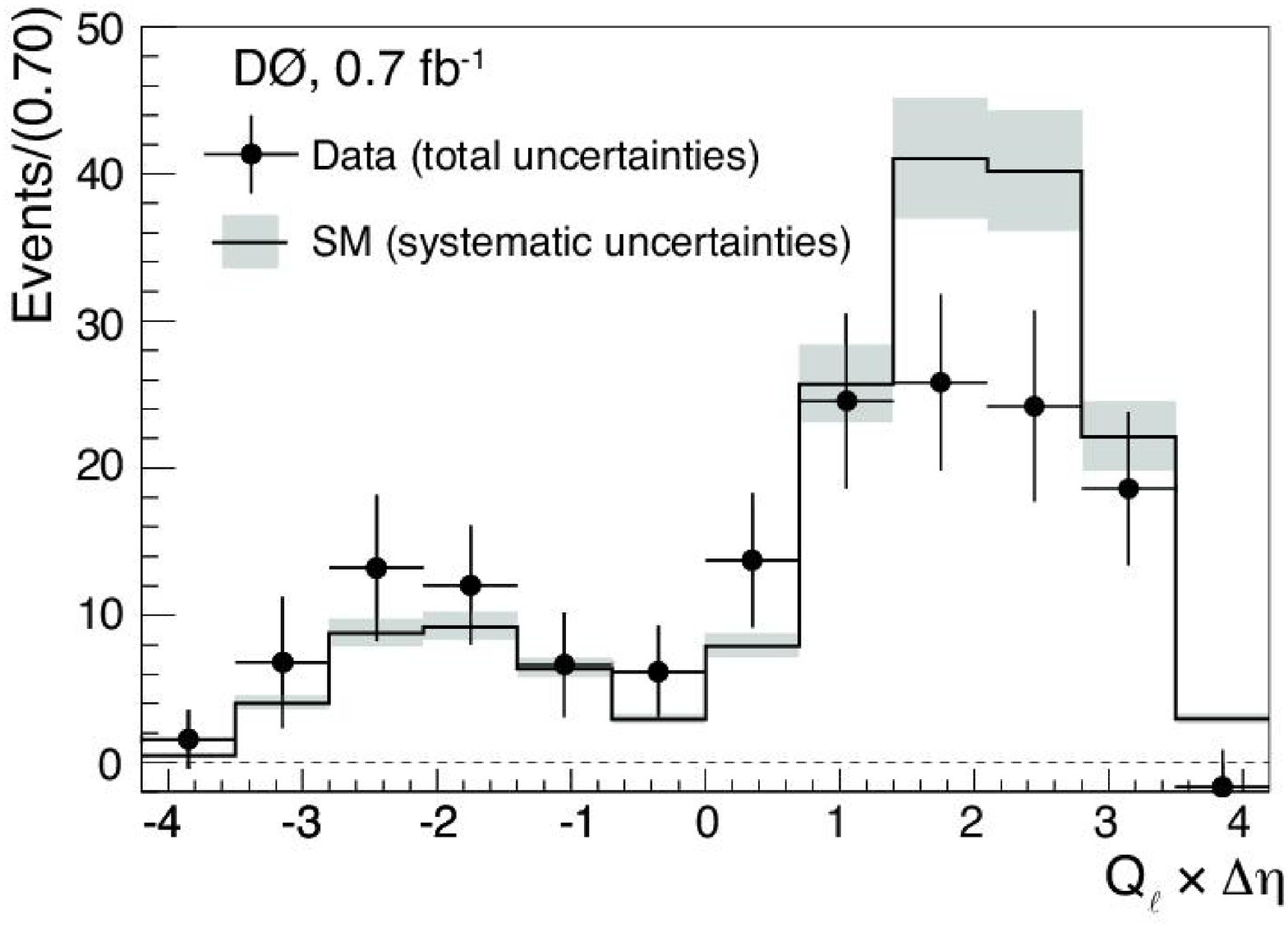}
\caption{Left: photon $E_T$ distribution for D\O\ $W\gamma$ candidate
events (points), and for the expected SM signal + background (open solid-line
histogram). The shaded histogram shows background contribution.
Dashed histogram corresponds to the non-SM coupling 
for $WW\gamma$~\cite{wg_raz_d0}.
Right: The background-subtracted charge-signed rapidity difference
for the D\O\ $W\gamma$ candidate
events (points), and for the expected SM signal + background (histogram)
~\cite{wg_raz_d0}.}
\label{fig:wg}
\end{figure*}

At leading order, the SM allows 
production of $p\bar{p}\to q\bar{q}' \to W \gamma$ via 
photon radiation off an incoming quark (initial state ration)
or directly through $WW\gamma$ vertex. These two 
production mechanisms involve three amplitudes where each alone violates
unitarity, but together interfere to give finite cross section.
This interference leads to radiation-amplitude zero (RAZ) in the angular
distribution of the photon. The RAZ manifests itself as a dip
in the charge-signed rapidity difference between the photon and the
charged  decay lepton from the $W$ boson, $Q_{\ell} \times \Delta \eta=
Q_{\ell}(\eta_{\gamma} - \eta_{\ell})$~\cite{raz_theory}.

Non-SM $WW\gamma$ couplings will give rise to
an increase in the $W\gamma$ production cross section over the SM
prediction, particularly for energetic photons. Anomalous TGCs can
also make RAZ dip more shallow or disappear entirely. 
Figure~\ref{fig:wg} (left) shows distribution of the photon $E_T$
in D\O\ $W\gamma$ candidates together with the expected
SM signal and background distributions. Example
of distribution for non-SM $W\gamma$ signal is also shown.
In order to set limits on anomalous TGCs, 
$W\gamma$ signal events are generated at various values
of TGCs. Observed photon $E_T$ spectrum in then compared
to the expected ones to determine the likelihood
that they represent the data. Obtained
one-dimensional 95\% C.L. limits by D\O\ are 
$-0.51<\Delta \kappa_{\gamma} < 0.51$ and 
$-0.12<\lambda_{\gamma} < 0.13$ for $\Lambda=2$~TeV~\cite{wg_raz_d0}.

Figure~\ref{fig:wg} (right) shows distribution of
the background-subtracted  $Q_{\ell} \times \Delta \eta$
for $W\gamma$ candidates in the D\O\ data together with
the SM expectation. The dip in the distribution at $Q_{\ell} \times \Delta \eta\simeq -0.3$
is clearly visible. In order to
evaluate the significance of the
observation, a set of anomalous coupling 
which provides a $Q_{\ell} \times \Delta \eta$
distribution that minimally exhibits no dip
is selected. This corresponds to $\kappa_{\gamma}=0$ and
$\lambda_{\gamma}=-1$ values of TGCs. For this set, probability to
observed the dip due to the random fluctuation
is estimated to be 4.5$\times 10^{-3}$  corresponding to 2.6~$\sigma$ Gaussian
significance. This constitutes the first indication of RAZ in 
$W\gamma$ production~\cite{wg_raz_d0}.

\begin{figure*}[t]
\centering
\includegraphics[width=75mm]{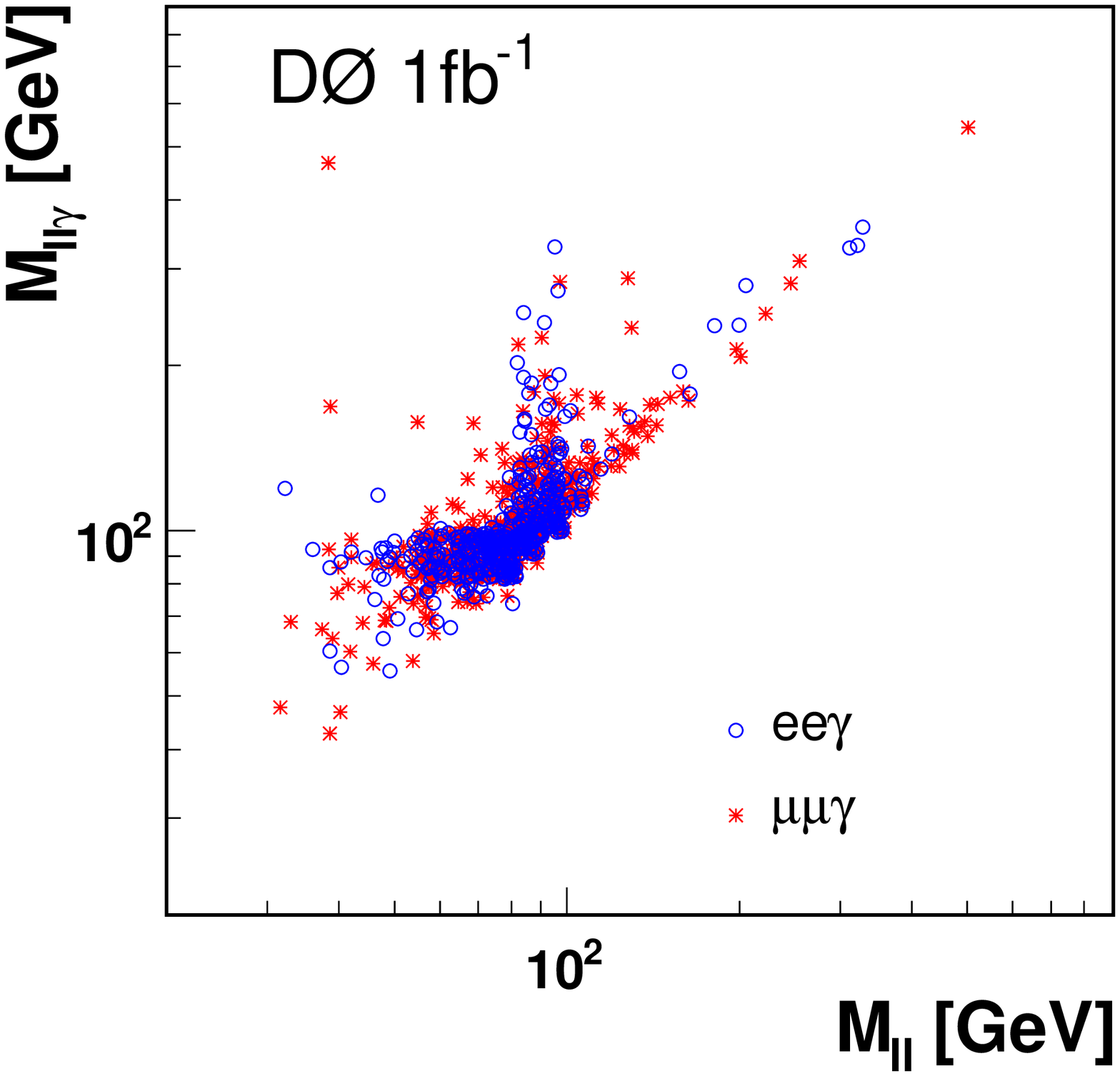}\hspace*{18mm}
\includegraphics[width=80mm]{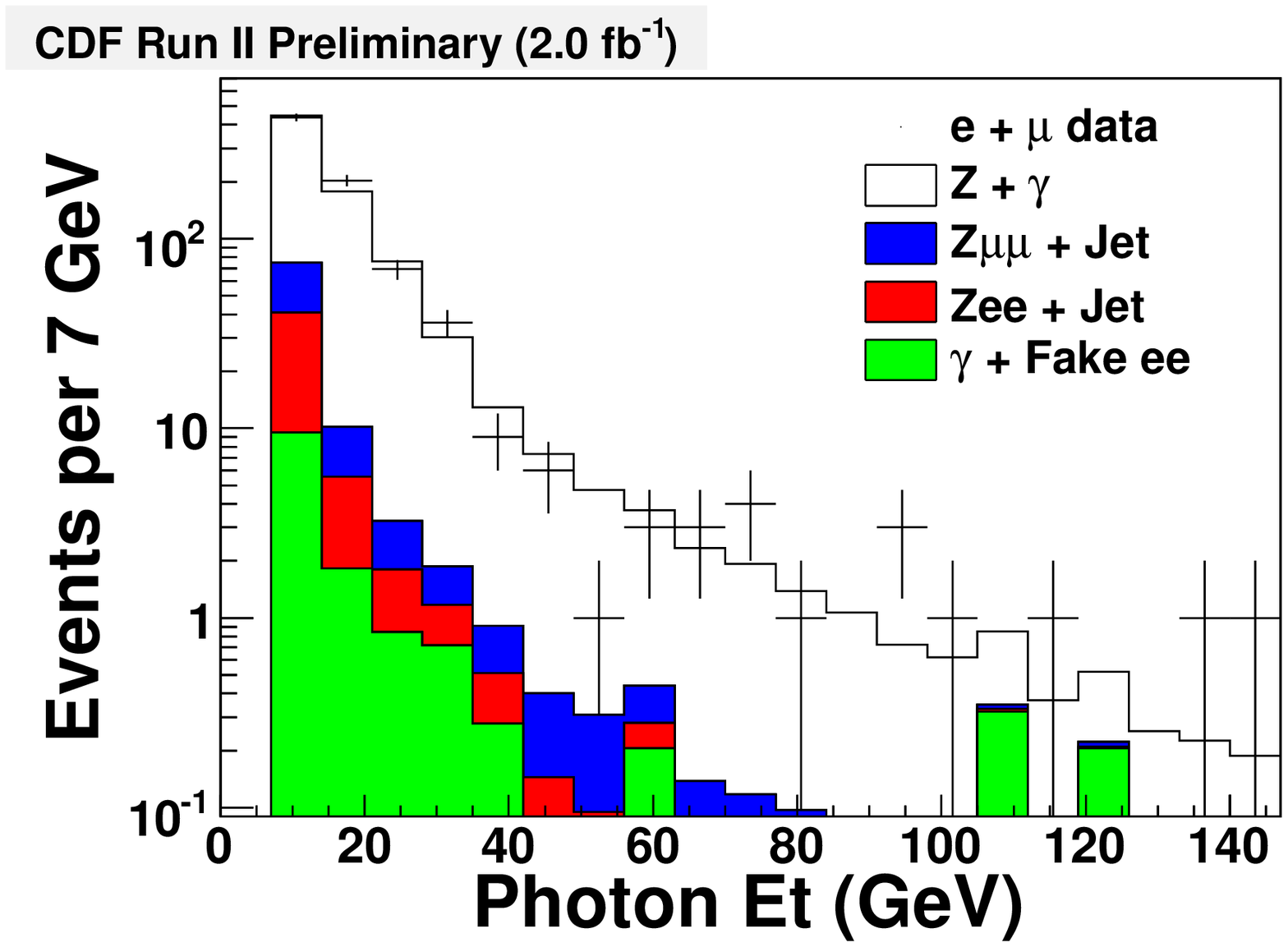}
\caption{Left: Dilepton + photon vs. dilepton mass in D\O\ $Z\gamma
\to ee\gamma/\mu\mu\gamma $
candidate events~\cite{zg_d0}. Masses of the candidates in 
electron channel are shown as open circles,
while those in the muon channel are shown as stars.
Right: Observed (points) and expected (histograms) distributions 
of photon $E_T$ in CDF $Z\gamma\to ee\gamma/\mu\mu\gamma$ analysis~\cite{zg_cdf}.} 
\label{fig:zg}
\end{figure*}

\section{$Z\gamma\to\ell\ell\gamma$ production}

Production of $p\bar{p}\to q\bar{q}\to Z\gamma$ events at the 
Tevatron is studied in $Z\gamma\to\ell^+\ell^-\gamma$ channel.
The signal sample
is selected by requiring a pair of either muon or electron,
and a photon. The photon can be produced by final state radiation
(FSR) off either charged leptons or one of the initial partons (ISR).
The main background process
is $Z+jet$ production where a jet is misidentified as
a photon. Both Tevatron experiments have measured
cross section times branching ratio for the 
 $p\bar{p}\to q\bar{q}\to Z\gamma$ production. 
Figure~\ref{fig:zg} (left) shows
 dilepton + photon vs. dilepton mass distribution
for D\O\ $Z\gamma \to ee(\mu\mu)\gamma $ candidate events.
The observed structure reflects three sub-processes:
the vertical band  with $M_{\ell\ell}\simeq M_Z$ and
$M_{\ell\ell\gamma}>M_Z$  corresponds to ISR production;
the horizontal band at 
$M_{\ell\ell\gamma}\simeq M_Z$ and $M_{\ell\ell}<M_Z$ 
corresponds to FSR events; the Drell-Yan events populate the diagonal band
with $M_{\ell\ell}\simeq M_{\ell\ell\gamma}$.

Using ${\cal{L}}=1$~fb$^{-1}$ of data, 
D\O\ has obtained $\sigma \times BR (Z\gamma\to \ell\ell\gamma)=
4.96\pm 0.30$~(stat+syst)~$\pm 0.30$~(lumi)~pb
for  $M(\ell\ell)>30$~GeV, $E_T^{\gamma}>$7~GeV 
and $dR(\ell\gamma)>$0.7~\cite{zg_d0}. The latter
requirement minimizes contribution from the FSR sub-process.
The measurement is in agreement with NLO SM expectation of
$\sigma \times BR (Z\gamma\to \ell\gamma]=4.74\pm0.22$~pb~\cite{zg_xs_theory}.
CDF has measured  $\sigma \times BR (Z\gamma\to \ell\ell\gamma)$
separately for ISR enriched ($M(\ell\ell\gamma) > 100$~GeV)
and FSR enriched ($M(\ell\ell\gamma) < 100$~GeV) 
productions obtaining  $\sigma \times BR (Z\gamma\to \ell\ell\gamma)=
1.2\pm 0.1$~(stat)$\pm 0.2$~(syst)$\pm 0.1$~(lumi)~pb,
and $\sigma \times BR (Z\gamma\to \ell\ell\gamma)=
3.4\pm 0.2$~(stat)$\pm 0.2$~(syst)$\pm 0.2$~(lumi)~pb, 
respectively~\cite{zg_cdf}.
The measurements use ${\cal{L}}=1.1(2.0)$~fb$^{-1}$  of data 
for $Z\gamma\to ee\gamma$ ($\mu\mu\gamma$) channel and are in good
agreement with the SM NLO theory calculations.

\begin{table}[t]
\begin{center}
\caption{One-dimensional 95 \% C.L. limits on anomalous 
neutral TGCs obtained by D\O\ and CDF in $Z\gamma$ analysis.
The limits correspond to form factor scale $\Lambda=1.2$~TeV.}
\vspace*{2mm}
\begin{tabular}{|c|c|}
\hline
 \textbf{D\O, ${\cal{L}}=$1~fb$^{-1}$~\cite{zg_d0}} & 
 \textbf{CDF, ${\cal{L}}=$1.1-2.0~fb$^{-1}$~\cite{zg_cdf}} \\ \hline
-0.085 $<h_3^{\gamma}<$0.084   & -0.084 $<h_3^{\gamma}<$0.084 \\  \hline
-0.0053$<h_4^{\gamma}<$0.0054  & -0.0047$<h_4^{\gamma}<$0.0047 \\  \hline
-0.083 $<h_3^Z<$0.082          & -0.083 $<h_3^Z<$0.083  \\  \hline
-0.0053$<h_4^Z<$0.0054         & -0.0047$<h_4^Z<$0.0047 \\  \hline
\end{tabular}
\label{table:zg_ac}
\end{center}
\end{table}

Most general effective Lagrangian that assumes Lorentz and gauge invariance,
has two CP-violating ($h_1^V$ and $h_2^V$) and
two CP-conserving ($h_3^V$ and $h_4^V$) parameters
for anomalous trilinear $ZV\gamma$ ($V=Z, \gamma$) 
couplings. Unitary is ensured by using form factor
parametrization $h_i^V=\frac{h_{i0}^V}{(1+\hat{s}/\Lambda^2)^n}$,
with $\Lambda$ being a form factor scale, $h_{i0}^V$
being the low-energy approximations of the couplings,
and n=3(4) for $h_{1,3}^V (h_{2,4}^V)$ ~\cite{zg_ac_theory}.
Parameters $h_i^V$ are all zero in the SM. Non-zero
$h_i^V$ couplings typically enhance $Z\gamma$ production
cross section, particularly at high values of photon $E_T$.
The $E_T$ distribution of the photon for CDF $Z\gamma$ candidate events, 
compared with the background and the SM $Z\gamma$ prediction is 
shown in Fig.~\ref{fig:zg} (right). To set limits
on anomalous $ZZ\gamma$ and $Z\gamma\gamma$ couplings,
photon $E_T$ distribution
in data is compared with the expected $E_T$ distribution from anomalous 
$Z\gamma$ production for a given set of $ZZ\gamma$ and $Z\gamma\gamma$
coupling values. Limits on anomalous TGCs obtained by D\O\ and CDF 
Collaborations are summarized in Table~\ref{table:zg_ac}. 
Obtained limits on $h_{40}^V$ are the most stringent to date.

\section{$ZZ\to\ell\ell\ell\ell$ and $ZZ\to \ell \ell \nu \nu$ productions}

The  NLO SM cross section for $p\bar{p}\to ZZ$ production
at $\sqrt{s}=1.96$~TeV is 
$\sigma(ZZ)=1.4\pm0.1$~pb~\cite{zz_xs_theory}. The process
has been studied in two decay modes
at the Tevatron: $ZZ\to \ell \ell \ell \ell$ and
$ZZ\to \ell \ell \nu \nu$ channels. The first mode is experimentally
very clean giving rise to events with four high-$p_T$ isolated leptons 
and very little hadronic activity. However, it also suffers
from low branching fraction of 4.5$\times$10$^{-3}$,
with total expected number of $ZZ\to \ell \ell \ell \ell$ events
being 6.3 per fb$^{-1}$. This is further reduced by
kinematic selection and lepton identification requirements.
Background to $ZZ\to \ell \ell \ell \ell$ signal arises
from $Z(\gamma)+jets$ and $t\bar{t}$ production processes and
is typically orders of magnitude smaller compared to the signal.
$ZZ\to \ell \ell \nu \nu$ channel has higher branching
fraction, but also higher background contamination
mainly from $WW$, $Z+jets$ and  $WZ$ productions
which can all produce events with two high-$p_T$ lepton
and missing $E_T$.

\begin{figure*}[t]
\centering
\includegraphics[width=79mm]{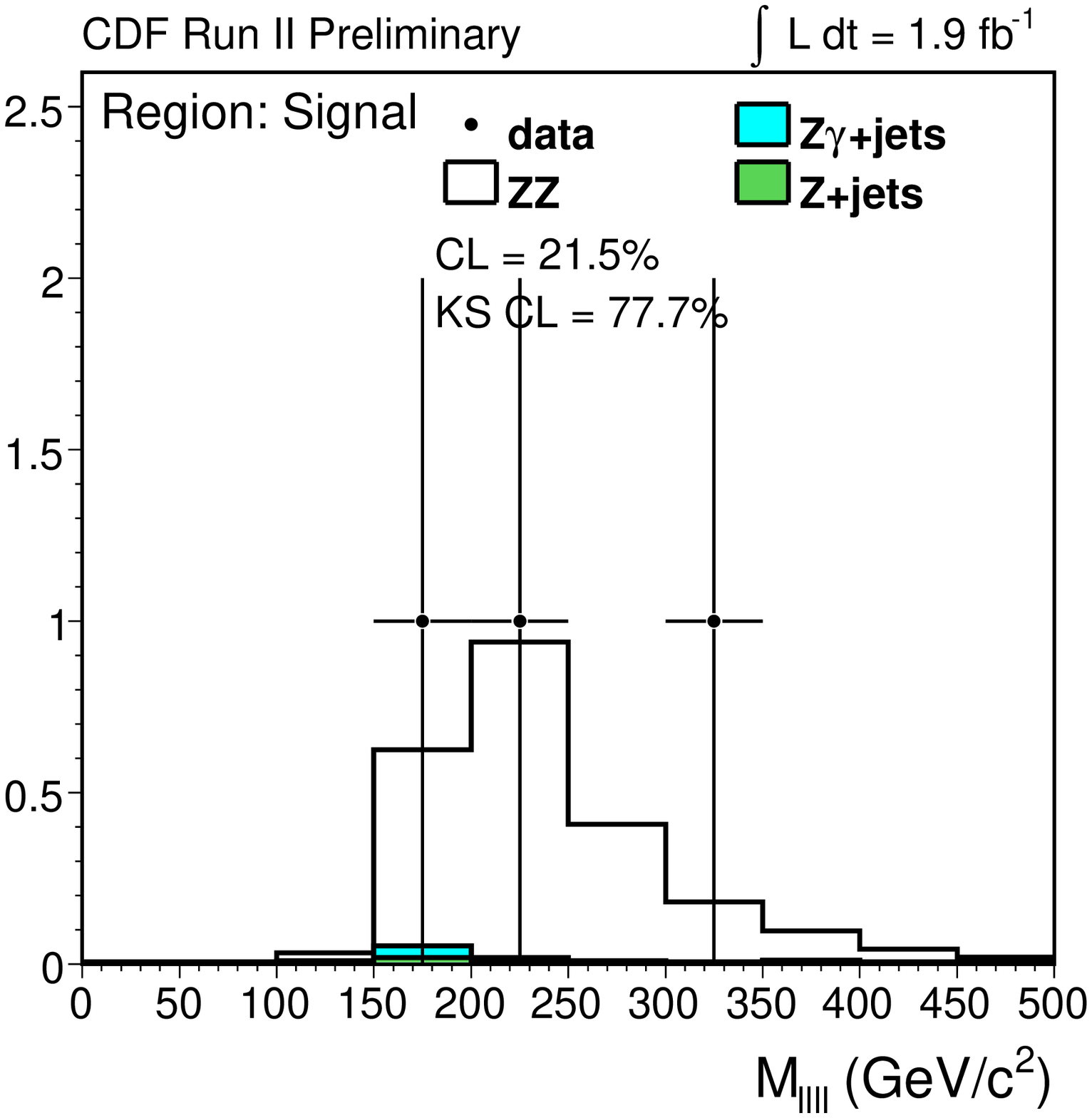}\hspace*{10mm}
\includegraphics[width=76mm]{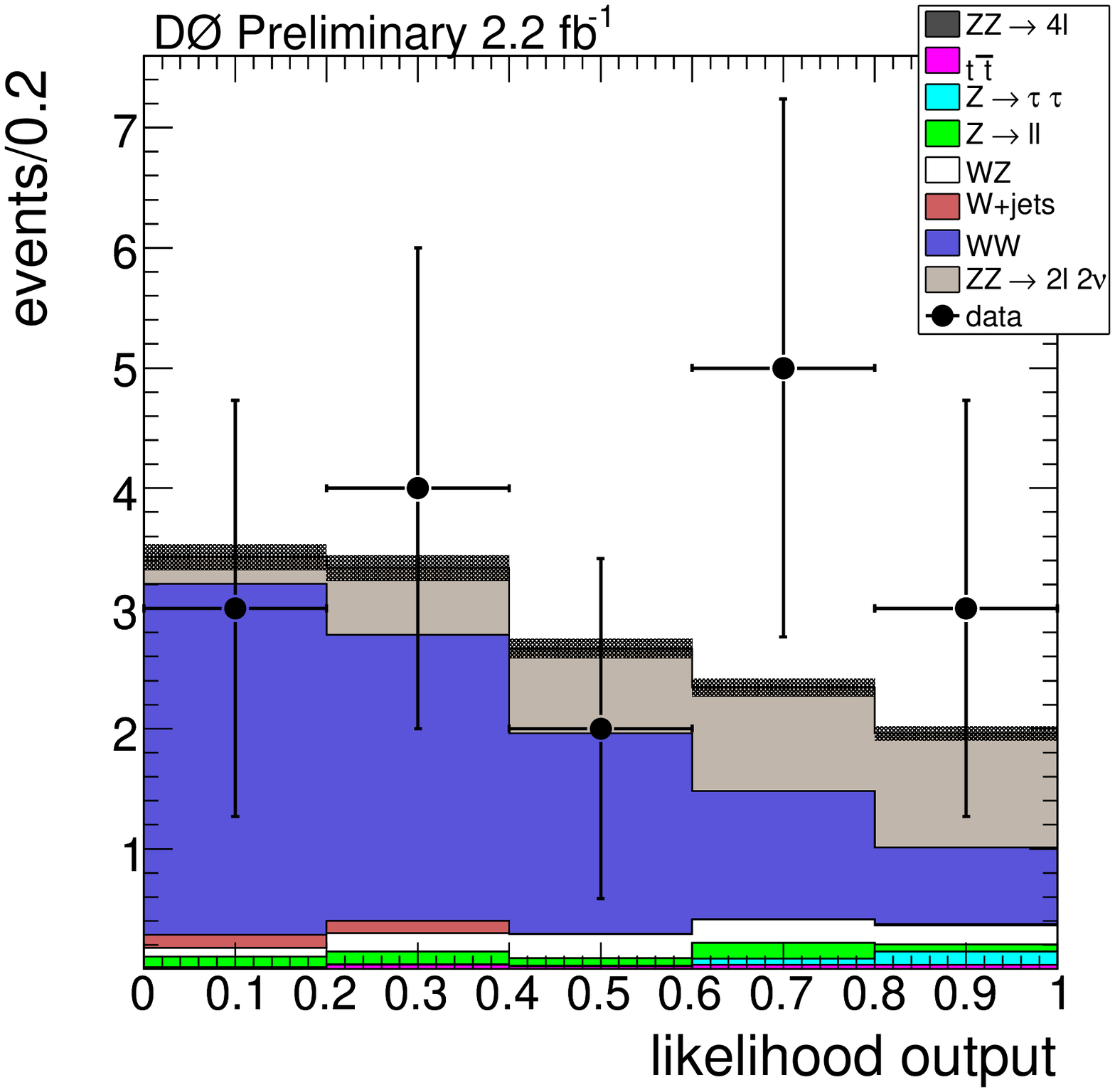}
\caption{Left: Observed (points) and expected distributions (histograms)
of four lepton invariant mass in $ZZ\to \ell \ell \ell\ell$ channel in CDF~\cite{zz_cdf}.
Right: Observed (points) and expected distributions (histograms)
of event likelihood discriminant in $ZZ\to\ell\ell\nu\nu$ channel in D\O~\cite{zz_llvv_d0}. 
} 
\label{fig:zz}
\end{figure*}

D\O\ and CDF experiments have both studied $ZZ$ production
in $ZZ\to \ell \ell \ell \ell$ and $ZZ\to \ell \ell \nu \nu$  channels.
In ${\cal{L}}=1.9$~fb$^{-1}$ of data, CDF has
observed 3 $ZZ\to \ell \ell \ell \ell$ candidates with
expected background of 0.096$^{+0.092}_{-0.063}$ events.
Figure~\ref{fig:zz} (left) shows four-lepton invariant
mass distribution for the three observed events,
as well as expected distributions for the background
and the signal.
For $ZZ\to \ell \ell \nu \nu$  channel, a leading order
calculations of the relative $ZZ$ and $WW$ event probabilities
is used to discriminate between signal and background.
Combination of $ZZ\to \ell \ell \ell \ell$ and
 $ZZ\to \ell \ell \nu \nu$ channels leads to
observation of excess over expected background
at the level of 4.4 $\sigma$ statistical significance.
The measured combined cross section of $\sigma(p\bar{p}\to
ZZ)=1.4^{+0.7}_{-0.6}$~(stat+syst)~pb~\cite{zz_cdf} is consistent
with the NLO SM expectation~\cite{zz_xs_theory}.

In ${\cal{L}}=1$~fb$^{-1}$ of data D\O\ has
observed one $ZZ\to \ell \ell \ell \ell$ candidate with
expected signal and background rates being
1.71$\pm$0.15 and $0.13\pm 0.03$ events, respectively.
This gives 95\% C.L. upper limit of 4.4~pb 
for $ZZ$ production cross section~\cite{zz_d0}.
For $ZZ\to \ell \ell \nu \nu$ channel,
several kinematic variables have been combined
in likelihood discriminant to achieve good signal-to-background
discrimination. Figure~\ref{fig:zz} (right) shows likelihood
distribution for data and for expected signal and background
in $ZZ\to \mu \mu \nu \nu$ channel. Obtained 
cross section value of  $\sigma(p\bar{p}\to
ZZ)=2.1\pm 2.1$~(stat)$\pm0.4$(syst)~pb~\cite{zz_llvv_d0} is consistent
with the NLO SM expectation~\cite{zz_xs_theory}.

Most general effective Lagrangian that assumes Lorentz and gauge invariance,
has two CP-violating ($f_4^V$) and
two CP-conserving ($f_5^V$) parameters
for anomalous trilinear $ZZV$ ($V=Z, \gamma$) 
couplings. Unitarity is ensured by using form factor
parametrization $f_i^V=\frac{f_{i0}^V}{(1+\hat{s}/\Lambda^2)^3}$,
with $\Lambda$ being a form factor scale and $f_{i0}^V$
being the low-energy approximations of the couplings~\cite{zz_ac_theory}.
$ZZZ$ and $ZZ\gamma$ vertices are all forbidden in the SM
at the tree level. Non-SM couplings typically increase $ZZ$ production
cross section. Using observed and expected number of 
events for various assumptions on $ZZZ$ and $ZZ\gamma$ 
coupling values, D\O\ has derived limits on anomalous TGCs ~\cite{zz_d0}.
One-dimensional 95\% C.L. limits are
-0.28$<f_{40}^Z<$0.28,
-0.26$<f_{40}^\gamma<$0.26,
-0.31$<f_{50}^Z<$0.29,
-0.30$<f_{50}^\gamma<$0.28. The limits are competitive to
those of the combined LEP experiments~\cite{lep}.

\section{summary}

Productions of $WW$, $WZ$, $W\gamma$, $Z\gamma$, and $ZZ$
have all been studied at the Tevatron. Measured production
cross sections for these processes are in agreement
with the SM expectations. The diboson productions
allow to directly probe Triple Gauge boson Couplings 
via observed event rates and kinematics.
With no indication for the deviation from the SM expectation,
limits are set on anomalous TGCs.
First indication of the peculiar feature, such as  
Radiation Amplitude Zero, predicted by SM 
for the $W\gamma$ production has also been observed.

\begin{acknowledgments}
The author wishes to thank HCP2008 
organizers for lively and constructive atmosphere at
the conference. 
\end{acknowledgments}

\end{document}